\RequirePackage{lineno}
\documentclass[aps,prc,twocolumn, floatfix]{revtex4}
\usepackage{graphicx}
\usepackage{dcolumn}
\usepackage{bm}
\usepackage{mhchem}
\usepackage{color}
\usepackage{multirow}

\begin{document}


\title{Examining the potential synthesis of new elements with $^{294}$Og}

\author{Shuai Zhang$^{1,2}$}
\author{Gao-Chan Yong$^{1,2}$}
\email[Corresponding author: ]{yonggaochan@impcas.ac.cn}
\author{J. L. Rodr\'{\i}guez-S\'{a}nchez$^{3,4}$}
\author{J. Cugnon$^{5}$}

\affiliation{
$^1$Institute of Modern Physics, Chinese Academy of Sciences, Lanzhou 730000, China\\
$^2$School of Nuclear Science and Technology, University of Chinese Academy of Sciences, Beijing 100049, China\\
$^3$CITENI, Campus Industrial de Ferrol, Universidade da Coru\~{n}a, E-15403 Ferrol, Spain\\
$^4$IGFAE, Universidad de Santiago de Compostela, E-15782 Santiago de Compostela, Spain\\
$^5$AGO department, University of Li\`{e}ge, all\'{e}e du 6 ao\^{u}t 19, b\^{a}t.~B5, B-4000 Li\`{e}ge, Belgium
}

\begin{abstract}

In the relentless pursuit of expanding the periodic table, the discovery of element 119 remains elusive, despite two decades of dedicated research efforts. The traditional fusion-evaporation approach, although fruitful in the past, now appears to be approaching its operational limits. This scenario sets the stage for considering innovative methodologies essential for further advancements in the field of superheavy elements. Here, we introduce a pioneering strategy aimed at synthesizing element 119 by adapting and extending the nuclear reaction processes previously successful in producing element $^{294}$Og. This involved the fusion of $^{48}$Ca and $^{249}$Cf. Building on this, our novel approach incorporates an additional reactive target --- specifically, hydrogen --- positioned strategically behind the $^{249}$Cf. This configuration is designed to facilitate an immediate secondary reaction of the nascent $^{294}$Og with hydrogen, potentially forging new pathways to element 119. Preliminary insights also suggest that employing isotopes like deuterium or helium-3 as targets may not only enhance the production rates of element 119 but might also pave the way for the synthesis of even heavier elements, extending up to elements 120 and 121. We delve into the technicalities and feasibility of employing a dual-target method using a $^{48}$Ca beam, exploring new horizons in the quest for the superheavy unknown.

\end{abstract}

\maketitle

%
Following the synthesis of element 118, the seventh row of the periodic table has been completed \cite{elem118}. In the field of superheavy element research, critical questions involve determining the heaviest nuclei that can exist and defining the limits of the periodic table \cite{sa19}. The production of superheavy elements has traditionally depended on heavy-ion fusion reactions near the Coulomb barrier \cite{fu1,fu2}. This method involves a nuclear fusion-evaporation reaction, using ion beams typically carrying lighter single or double magic nuclei, known as projectiles, directed at thin targets composed of heavier single or double magic nuclei \cite{hof2015}. Due to the extremely low probability of creating superheavy nuclei, a potent heavy-ion beam, a stable heavy actinide target, and advanced target technology are required \cite{hof2015,cc21}. Moreover, theoretical studies highlight the uniqueness of superheavy nuclei due to their significant charges, which trigger quantum shell effects and cause substantial Coulomb frustration within the nucleus, challenging current many-body theoretical frameworks \cite{np18}. Therefore, the choices of projectile and target combinations for synthesizing element 119 and subsequent elements are fraught with significant theoretical uncertainties.

Considering the practicality of a stable isotope beam and an actinide target in terms of radiation safety and chemical handling, the RIKEN Nishina Center (RNC) opted to use $^{51}$V as the projectile and $^{248}$Cm as the target. The objective of this pairing is to produce element Z = 119 through a hot fusion reaction of $^{51}$V + $^{248}$Cm at a beam energy of approximately 6 MeV/nucleon in the laboratory setting \cite{jap2022}. The peak theoretical cross section for the creation of element 119 in this reaction is projected to be 0.005 pb \cite{ppt2024}. Over the past two decades, extensive studies have been conducted on the synthesis of superheavy elements such as 119, 120, and 122 using various beam-target combinations \cite{shn1,shn2,shn3,shn4,hof2015}. Nevertheless, theoretical research suggests that the potential cross sections for producing these superheavy nuclei could range between 10$^{-1}$ and 10$^{-6}$ pb, highlighting considerable uncertainty in synthesis \cite{unc1,unc2,unc3}. Thus, there is a clear necessity to investigate new methods for synthesizing superheavy elements.

Currently, the lifespan of the $^{295}$Og isotope is 1000 times greater than that of the $^{294}$Og isotope \cite{cpc, epja}. While there are no more stable isotopes for element 118 (Og), numerous studies highlight the notably high fission barriers and substantial neutron and proton separation energies for several $_{118}$Og isotopes \cite{PRC91p, AP401p}. This suggests the possibility of finding longer-lived $_{118}$Og isotopes, particularly if alpha decay can be mitigated. Additionally, utilizing unstable atomic nuclei as projectiles through methods such as dual-target measurements and inverse kinematic collisions offers new avenues \cite{dual23,dual24}. These approaches could facilitate the synthesis of new, heavier elements by enabling reactions involving other atomic nuclei and unstable $_{118}$Og nuclei.

In this Perspective, we begin our discussion by addressing the synthesis of element 118, known as $^{294}$Og, via the proven technique involving the reaction of $^{48}$Ca and $^{249}$Cf. Adopting this method, we then employ the dual-target technique to promote reactions between $^{294}$Og and light particles such as protons, deuterons, helium-3, lithium-7, or other atomic nuclei, aiming to synthesize element 119 or higher. Utilizing a two-stage process model \cite{twostep}, our research initially implements a cascade collision model with a predetermined potential well to examine the interaction between the light particle and the target nucleus, including particle emission and the formation of the hot residual nucleus. This is followed by the application of a deexcitation model to meticulously analyze the final products emitted from the hot residual nucleus.

%
\begin{figure}[t]
\centering
\vspace{0.2cm}
\includegraphics[width=0.5\textwidth]{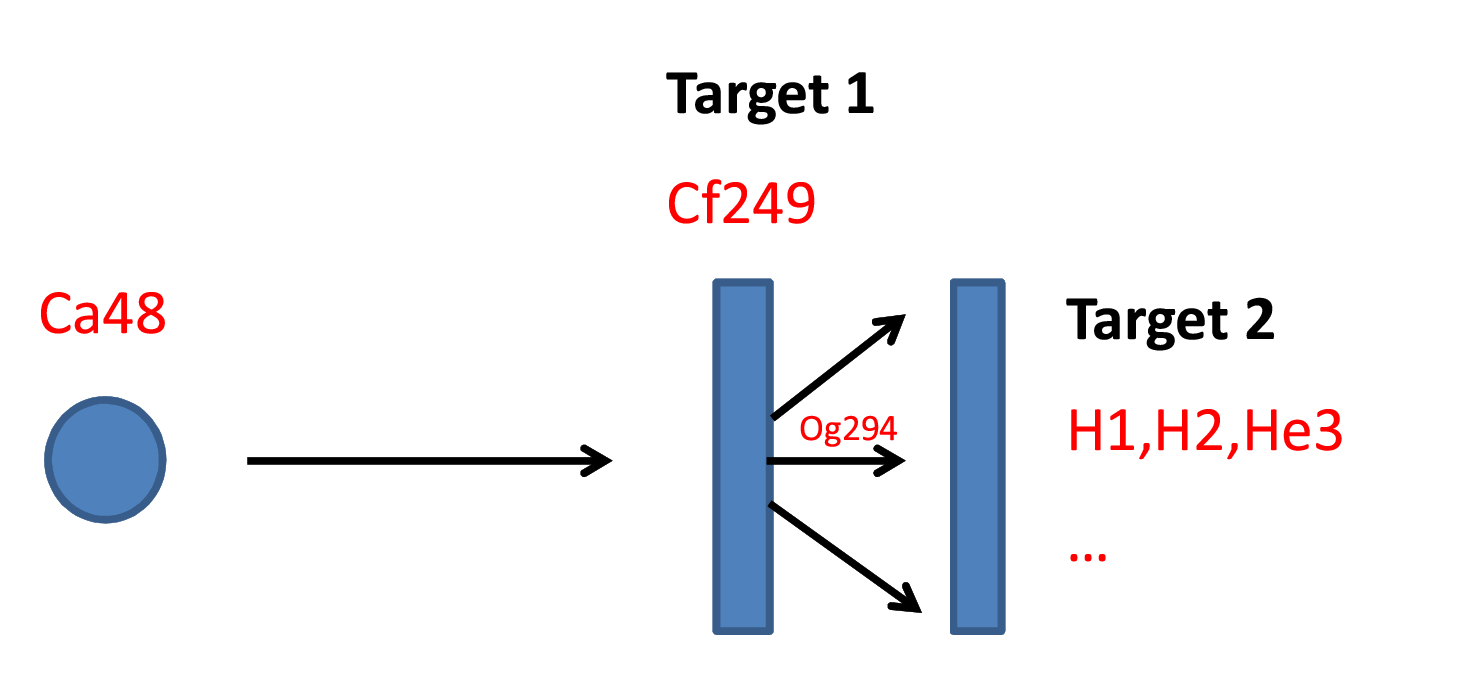}
\caption{Schematic diagram of synthesizing new element 119 based on the dual-target technique. A $^{48}$Ca beam reacts with $^{249}$Cf to synthesize element 118 ($^{294}$Og); then $^{294}$Og reacts with a hydrogen (or deuterium, helium-3, etc.) target to synthesize element 119.} \label{diagram}
\vspace{0.25cm}
\end{figure}
The Li\`{e}ge Intranuclear Cascade (INCL) model, extensively utilized in various studies \cite{incl2002, incl13, incl14}, depicts the interactive dynamics between projectiles (including nucleons, pions, hyperons, and light nuclei) and target nuclei. Rooted in the principles of classical physics, the model adeptly incorporates quantum-mechanical effects to replicate the dynamics and pre-collision conditions. A pivotal aspect of this model is the utilization of nucleon potentials, influenced by energy and isospin, along with a constant isospin-sensitive hyperon potential based on the Woods-Saxon density distributions \cite{incl13, incl2017, incl2018}. The model approaches nuclear interactions through a series of relativistic binary hadron-hadron collisions and meticulously follows the temporal development of hadrons' positions and momenta. Enhancements to the INCL model include a novel framework for generating pion mesons and strange particles \cite{incl11pion, incl18strange}. The latest iteration, INCL++6.33, now also supports the generation of superheavy elements up to element 119 and beyond. Adhering to various conservation laws such as baryon number, charge, energy, momentum, and angular momentum, the model forecasts the emergence of hot superheavy remnants and classifies these remnants by atomic and mass numbers, excitation energy, and angular momentum.

Meanwhile, the ABLA deexcitation model offers a comprehensive analysis of the various phenomena involved in post-collision deexcitation \cite{abla07, incl22strange}. This dynamic framework deciphers the deexcitation of the thermalized system by considering simultaneous break-ups, particle emissions, and fission events. It describes simultaneous break-up as a thermal instability-induced disintegration of the heated nucleus into multiple fragments. The theoretical basis for particle evaporation follows the Wei?kopf-Ewing approach, incorporating dynamic factors in calculating the fission decay width. The model also extensively elaborates on the emission process, covering not only neutrons, light charged particles, and gamma rays but also intermediate-mass fragments. The prerequisites for the ABLA model are predefined, consistent across all systems and energies, which bolsters its predictive reliability \cite{abla07}.

%
\begin{figure}[t]
\centering
\vspace{-0.2cm}
\includegraphics[width=0.485\textwidth]{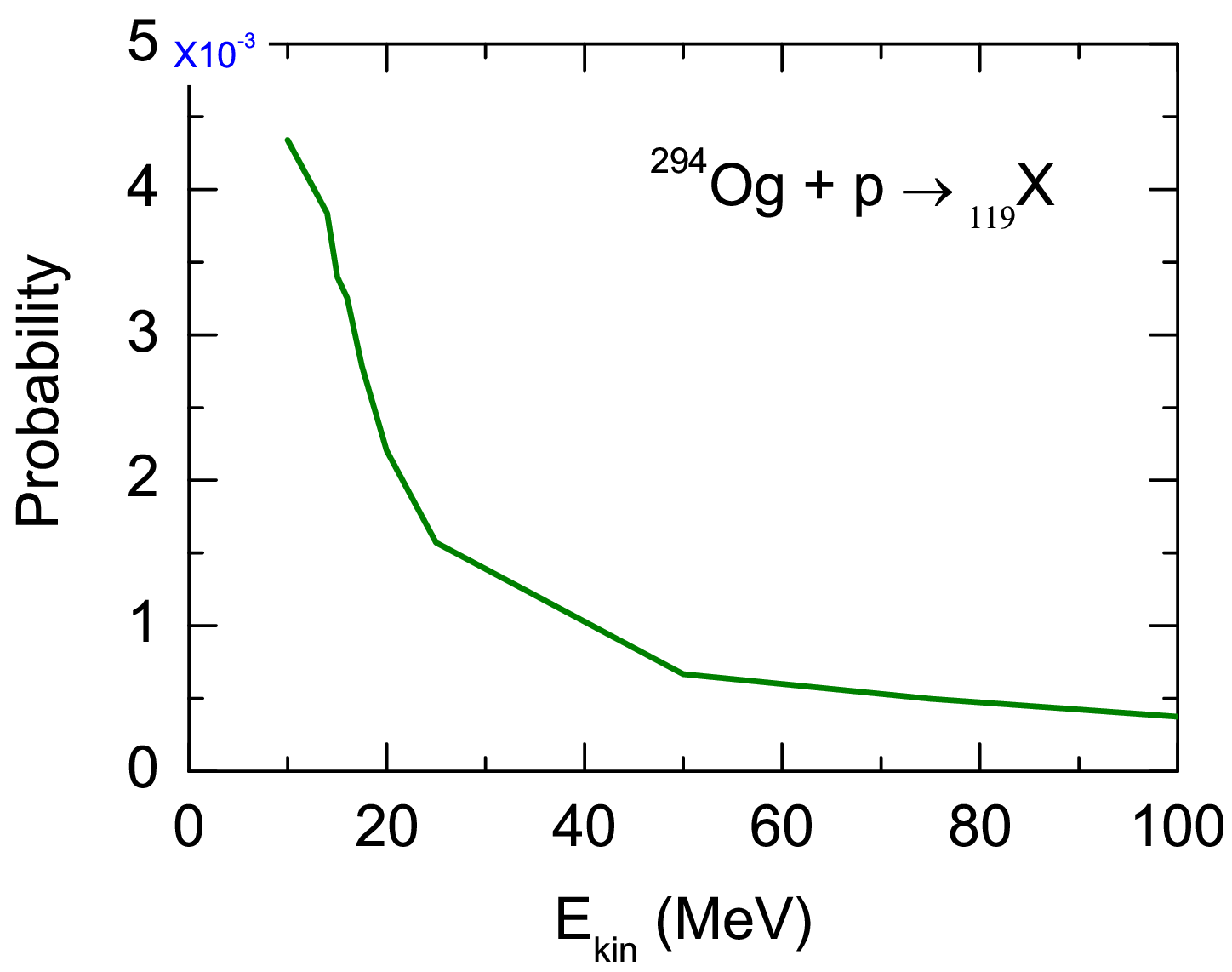}
\caption{Probability of synthesizing element 119 through the reaction of $^{294}$Og with a hydrogen target as a function of beam energy. The calculation is based on the INCL6.33+ABLA with the inverse kinematics method (where the beam energy is equal to the proton kinetic energy).} \label{prob}
\vspace{0.25cm}
\end{figure}
In this research, we introduce a new approach for synthesizing superheavy elements by initiating nuclear reactions between the heaviest currently known superheavy element, $^{294}$Og, and light particles to forge elements 119 and possibly element 120. Fig.~\ref{diagram} depicts the concept for a synthesis device specifically designed for superheavy elements such as element 119. Initially, element 118 ($^{294}$Og), the heaviest available element, is produced. Subsequently, this element 118 reacts with protons, deuterium, or other light nuclei to generate element 119. Considering the brief half-life of the intermediate $^{294}$Og, a dual-target technique is adopted to allow for the immediate interaction of freshly produced $^{294}$Og with light nuclei targets, thereby synthesizing element 119. Since the production of $^{294}$Og has been already experimentally proven and considered viable, our investigation concentrates exclusively on the reactions involving light nuclei targets and $^{294}$Og to synthesize element 119. As the target materials are gradually depleted, a rotational mechanism for the light nuclei targets could be utilized to refresh the depleted materials. Moreover, in this setup, subsequent targets might also involve heavier atomic nuclei, such as the doubly magic nucleus $^{16}$O and heavier isotopes of Ca, Ti, and so forth, potentially leading to the creation of superheavy elements beyond element 122. Nonetheless, overcoming Coulomb repulsion poses a challenge in these reactions. This issue can be tackled by integrating a superconducting linear accelerator between the targets.

From a historical perspective, $^{48}$Ca and $^{249}$Cf were used to synthesize $^{294}$Og, achieving a kinetic energy of 0.13 MeV/nucleon \cite{cross2942}. Implementing a superconducting linear accelerator with a gradient of 50 MeV per meter \cite{hey2020} and extending its length to 72 meters (considering the 0.13 MeV kinetic energy per nucleon and a 1 ms half-life, allowing $^{294}$Og to travel approximately 5000 meters before decaying, with the necessity to account for target thickness and energy loss, doubling the length of the accelerator might be necessary), allows $^{294}$Og to surpass the Coulomb barrier and subsequently react with protons or deuterons to create isotope 119.

Fig.~\ref{prob} illustrates the probability of producing element 119 through the reaction between protons and $^{294}$Og. It is evident that an increase in beam energy correlates with a decrease in the production rate of element 119. At a beam energy of 10 MeV, the production rate stands at roughly 0.45\%. Thus, to enhance the reaction rate, one should consider either increasing the beam intensity of $^{48}$Ca or augmenting the density of the target material. Additionally, Fig.~\ref{prob} highlights that lower beam energies significantly boost the likelihood of synthesizing the new element 119, consistent with the low kinetic energy per nucleon noted in $^{294}$Og's production, thereby favoring the synthesis of the new element 119.

\begin{figure}[t]
\centering
\vspace{0.2cm}
\includegraphics[width=0.48\textwidth]{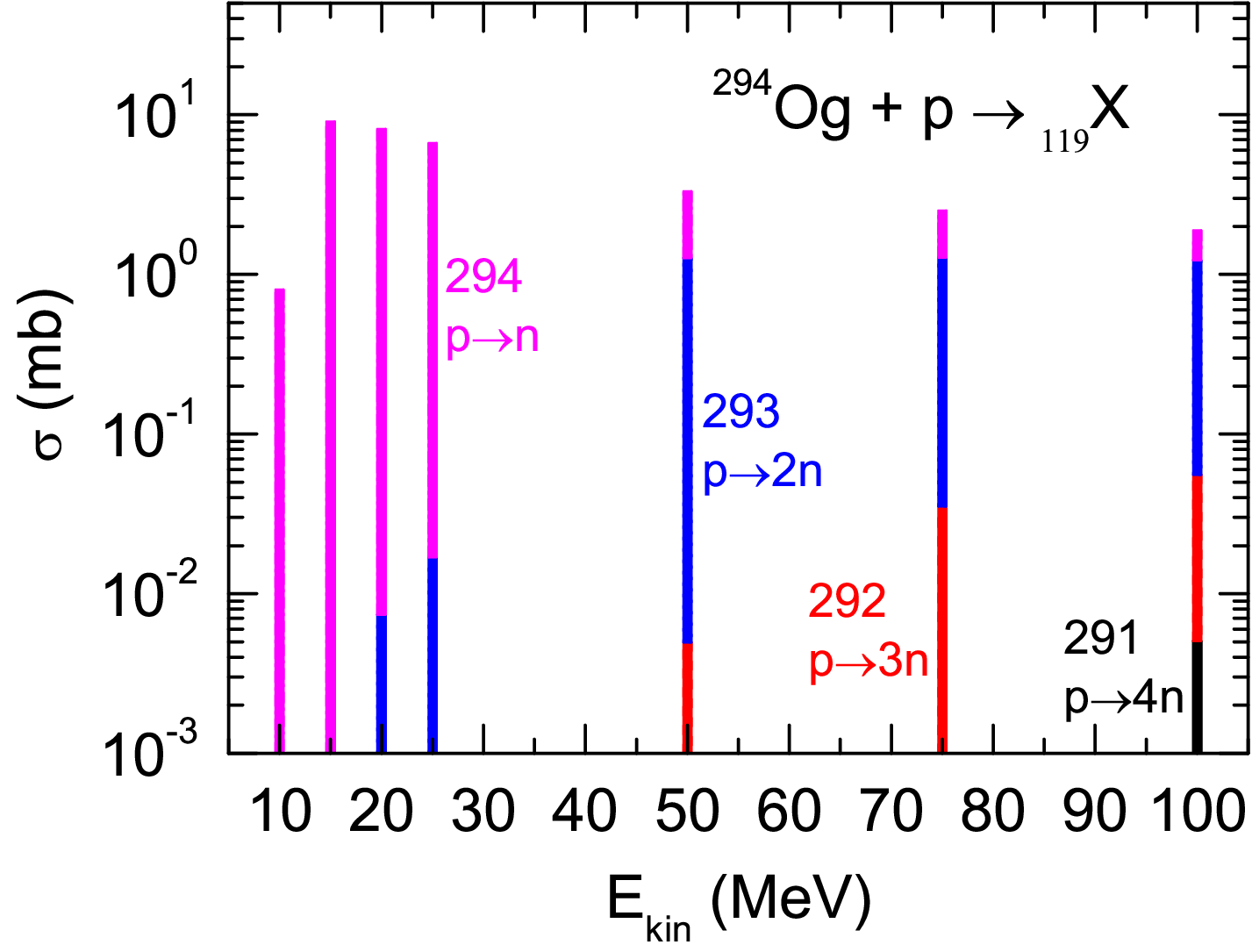}
\caption{Production cross sections of different isotopes of element 119 as a function of beam energy in the reaction of $^{294}$Og with a hydrogen target for the synthesis of element 119.} \label{xsection}
\vspace{0.25cm}
\end{figure}
\begin{figure}[t]
\centering
\vspace{-0.2cm}
\includegraphics[width=0.48\textwidth]{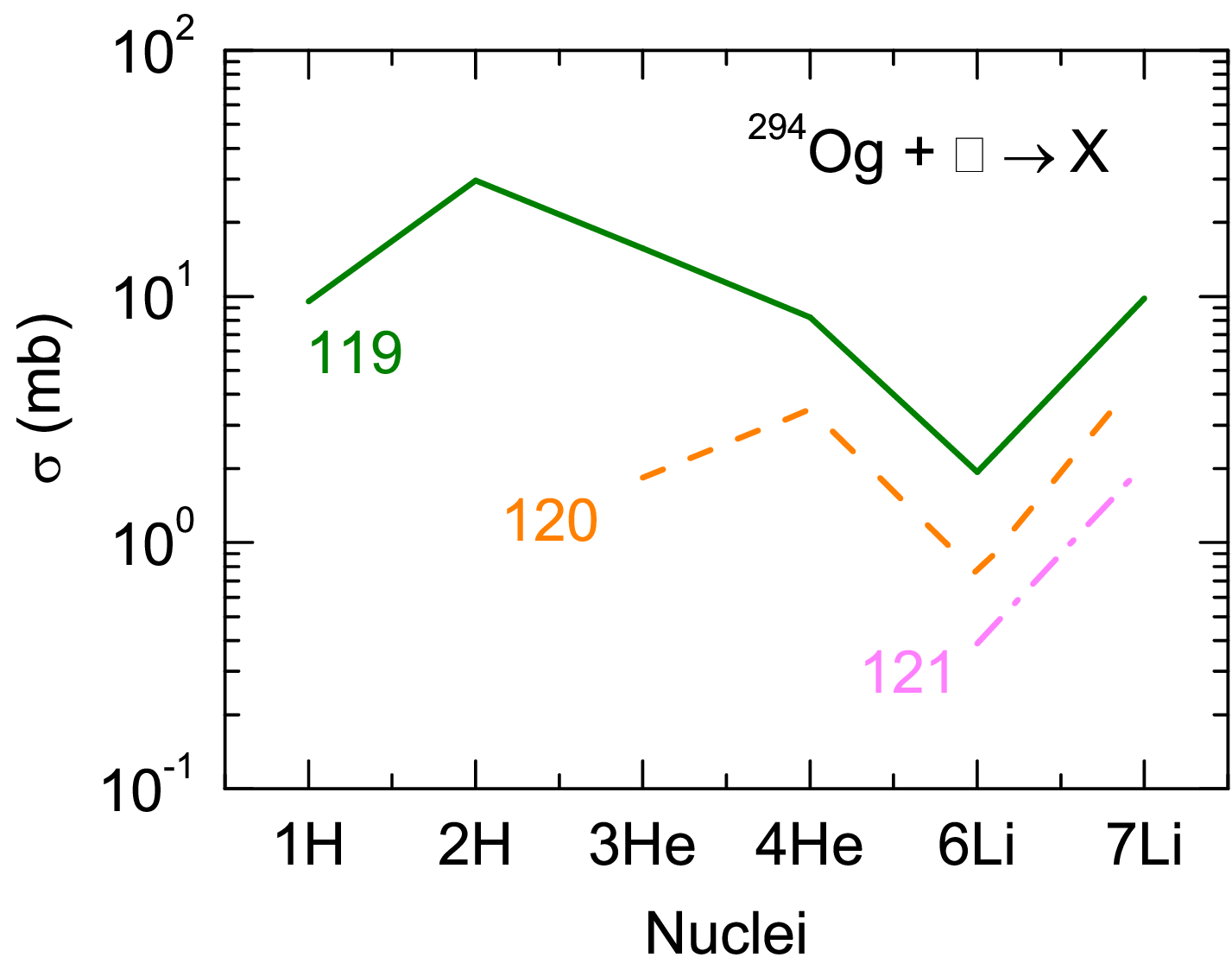}
\caption{Production cross sections of element 119 and 120, 121 with different target materials in reactions with $^{294}$Og and different targets (hydrogen target and deuterium target with a beam kinetic energy of 16 MeV; helium target with a beam kinetic energy of 48 MeV; lithium target with a beam kinetic energy of 96 MeV).} \label{xtarg}
\vspace{0.25cm}
\end{figure}
To delve deeper into the overall production cross section of element 119 and its various isotopes, as well as their variations with beam energy, we present Fig.~\ref{xsection}. Initially, it is observed that the total cross section for producing element 119 through the $^{294}$Og + p reaction peaks at approximately 16 MeV and subsequently declines as beam energy increases. At lower energies, the production predominantly leads to isotopes with mass numbers 294 and 293. This trend is expected because lower beam energies generate minimal excitation energies. Consequently, the $^{294}$Og + p reaction tends to produce element 119 mainly via the evaporation residue reaction mode, where evaporating one neutron results in mass number 294 and two neutrons yield 293. In contrast, at higher beam energies, there tends to be a higher neutron emission rate, which reduces the mass numbers of the resulting isotopes of element 119 (i.e., producing more isotopes with mass numbers 293, 292, or even 291). Furthermore, it is evident that lower beam energies result in a relatively pure production of isotope 294, whereas higher energies yield a wider array of isotopes. This analytical insight is crucial for understanding the synthesis dynamics of element 119. However, it should be noted that these cross sections are theoretical and may not directly reflect the actual synthesis rates of element 119.

In the proposed system, starting with $^{48}$Ca + $^{249}$Cf $\rightarrow$ $^{294}$Og and subsequent reaction of $^{294}$Og + p as depicted in Fig.~\ref{diagram}, the estimated number of element 119 isotopes synthesized can be calculated using the formula N$_{event}$ = N$_{proj}$$\times(1-e^{-\sigma/T_{thick}})$ \cite{est2014}, where $\sigma$ represents the reaction cross section and $T_{thick}$ denotes the target's particle or ion density per unit surface area. This density is derived from dividing the target's molar mass by the number of target nuclei per unit area. Assuming a $^{48}$Ca beam intensity of 9.25$\times$10$^{13}$ particles per second \cite{HIAF} --- about four times that of RIKEN's superheavy element beam \cite{jap2022} --- and a $^{249}$Cf target thickness of 0.34 mg/cm$^{2}$ \cite{cross294,cross2942} with a liquid hydrogen target of about 100 mg/cm$^{2}$, a reaction with a cross section of 10 mb for $^{294}$Og + p as shown in Fig.~\ref{xsection} will produce 2.22$\times$10$^{-8}$ isotopes of element 119 each second, leading to the synthesis of approximately one isotope of element 119 every 520 days. Utilizing deuterium instead, with a reaction cross section of about 30 mb (see Fig.~\ref{xtarg} below), the time to synthesize one isotope of element 119 is reduced to about 173 days. However, considering the energy loss of $^{294}$Og within a target of certain thickness, the actual kinetic energy per nucleon when $^{294}$Og interacts with hydrogen or deuterium targets may be higher, potentially halving the average reaction rate for element 119 production. Consequently, the number of days realistically needed could double. Furthermore, unlike traditional fusion-evaporation models, our INCL + ABLA model avoids a fusion phase when using a deuterium target, reducing model uncertainty. This model also incorporates more detailed effects such as nucleon-nucleon short-range correlations (SRC) \cite{src2014}, enhancing the reliability of our calculations. Therefore, the synthesis of element 119 as designed here is indeed feasible.

To explore diverse experimental setups, we assessed the production of elements 119, 120, and 121 by substituting the hydrogen target with alternative materials. Fig.~\ref{xtarg} illustrates the variation in the production cross section of element 119 when the target is changed sequentially from hydrogen to deuterium, helium-3, helium-4, lithium-7, and lithium-8. For each target element, we selected beam energies of 16, 48, and 96 MeV to ensure that all neutron-deficient isotopic targets are subjected to the same kinetic energy per nucleon. Since the presence of additional neutrons does not influence Coulomb repulsion within the reaction, neutron-rich targets are exposed to identical beam energy settings as their neutron-deficient counterparts. Changes in lighter targets reveal notable fluctuations in the cross section for producing element 119, often varying by several times. Employing targets like hydrogen, deuterium, helium-3, or lithium-7 in reactions with $^{294}$Og appears to facilitate the synthesis of new element 119 more readily. Additionally, as shown in Fig.~\ref{xtarg}, reactions involving $^{294}$Og and lighter targets could yield even heavier superheavy elements, such as elements 120 and 121. Moderately heavier targets such as helium and lithium are effective in producing these heavier superheavy elements. In contrast to element 119, the production cross sections for elements 120 and 121 are progressively lower, though the decrease is not by orders of magnitude.

The ongoing exploratory research into novel methods for synthesizing superheavy elements 119, 120, and potentially 121, as noted earlier, indicates that the likelihood of producing new elements using $^{294}$Og and lighter targets remains below approximately one percent. To increase this probability, a more intense $^{48}$Ca beam and denser target materials are necessary. Nonetheless, since the synthesis of $^{294}$Og through the reaction of $^{48}$Ca with $^{249}$Cf is well-documented, and the INCL+ABLA model has demonstrated considerable success in simulating reactions between heavy and light nuclei to generate new isotopes, our current exploratory research utilizing dual-target technology for synthesizing new superheavy elements 119 and 120 should offer valuable insights.

%
In summary, we have devised an experiment using a dual-target setup aimed at synthesizing new superheavy elements, starting with element 119. Specifically, building on the established synthesis of element 118, $^{294}$Og, via the reaction of $^{48}$Ca with $^{249}$Cf, we seek to extend this process to its neighboring elements. Considering the brief half-life of $^{294}$Og, we suggest that immediately after its formation, $^{294}$Og should interact with a target comprised of light nuclei such as hydrogen, deuterium, helium, or lithium to potentially produce element 119 or beyond. This approach utilizes dual-target technology, where a light nuclei target (designated "light" due to minimal Coulomb repulsion as $^{294}$Og possesses low kinetic energy) is positioned behind the $^{249}$Cf target. The reaction sequence is initiated by $^{48}$Ca, as follows: $^{48}$Ca + $^{249}$Cf $\rightarrow$ $^{294}$Og; then $^{294}$Og + p (or other light nuclei) $\rightarrow$ element 119, etc. Given that the reaction efficiency of $^{294}$Og + p leading to element 119 is below one percent, a high-intensity $^{48}$Ca beam is essential. We anticipate that this exploratory research will offer valuable insights for future efforts to synthesize new, heavier elements.

%
The author G.C.Y. thanks D. Mancusi and J.-C. David for helpful communications.
This work is partly supported by the National Natural Science Foundation of China under Grant Nos. 12275322, 12335008 and CAS Project for Young Scientists in Basic Research YSBR-088. J.L.R.-S. is thankful for the support provided by the Regional Government of Galicia under the program ``Proyectos de excelencia'' Grant No. ED431F-2023/43, and by the ``Ram\'{o}n y Cajal'' program under Grant No. RYC2021-031989-I funded by MCIN/AEI/10.13039/501100011033 and ``European Union NextGenerationEU/PRTR''.

\end{document}